\documentclass[onecolumn,showpacs]{revtex4}
\usepackage{graphicx}
\usepackage{dcolumn}
\usepackage{bm}
\input epsf
\begin{document}
\title{\Large  INTERACTING THREE FLUID SYSTEM AND THERMODYNAMICS OF THE UNIVERSE BOUNDED BY THE EVENT HORIZON}
\author{\bf~Nairwita~Mazumder\footnote{nairwita15@gmail.com},Ritabrata~Biswas\footnote{biswas.ritabrata@gmail.com}, Subenoy~Chakraborty\footnote{schakraborty@math.jdvu.ac.in}.}
\affiliation{$^1$Department of Mathematics,~Jadavpur
University,~Kolkata-32, India.}
\date{\today}
\begin{abstract}

The work deals with the thermodynamics of the universe bounded by
the event horizon. The matter in the universe has three
constituents namely dark energy, dark matter and radiation in
nature and interaction between then is assumed. The variation of
entropy of the surface of the horizon is obtained from unified
first law while matter entropy variation is calculated from the
Gibbss' law. Finally, validity of the generalized second law of
thermodynamics is examined and conclusions are written point wise.
\end{abstract}
 \maketitle
\section{\normalsize\bf{Introduction}}

At present it is strongly believed that our universe is
experiencing an accelerated expansion.The various cosmologicals
observations (for examples data collected from SNeIa [1], WMAP
[2], SDSS [3] and X-ray [4]) suggest that the acceleration is
driven by a missing energy density with negative pressure, known
as dark energy (DE). Although cosmological constant is a possible
solution for accelerated expansion but it is normally discarded
due to fine-tuning problem [5]. There are various candidates for
DE model namely variable Cosmological Constant[6], scalar field or
quintessence field [7], phantom field [8] (a scalar field with
negative kinetic term) or quintom field [9] (a unified model of
quintessence and phantom). From the effective quantum field theory
and gravitational effect some speculation about the nature of DE
is made and is known as holographic dark energy model (HDE) [10].
(One may note that from the recently proposed Horava gravity [11]
the DE model may have strong quantum gravitational back ground
[12]). Although, these DE models are in satisfactorily agreement
with observational evidences but a new type of 'coincidence'
problem [5] has come into picture- the density of vacuum energy
and that of dark matter (DM) are of the same order although they
have quite distinct energy scale during expansion era. A possible
way out of this problem is to consider interaction between DE and
DM. Further, from the analysis of the cosmic microwave back ground
radiation, our universe still have some remnant of this back
ground radiation. So it is quite natural
to consider DE interacting with both DM and radiation.\\

In the present work, we perform a thermodynamical analysis of the
universe bounded by the event horizon and the matter is chosen as
above i.e. DE interacting with DM and radiation. Thermodynamical
study of the universe bounded by the apparent horizon is common in
the literature as they form a bekenstein system [13] while any
definite character is still unknown for event horizon. Basically
we examine the validity of the generalized second law of
thermodynamics on the event horizon and the required constraints
are analyzed. Also a comparison with earlier results
are attempted.\\

\section{\normalsize\bf{Formulation of the model:}}

Let us consider our universe to be homogeneous and isotropic FRW
model bounded by the event horizon. The universe is assumed to be
filled up with DE interacting with DM and radiation. The space
time geometry is described by the line element

\begin{equation}
ds^{2}=-dt^{2}+a^{2}(t)\left[\frac{dr^{2}}{1-kr^{2}}+r^{2}d\Omega^{2}\right]
\end{equation}

$$=h_{ab}dx^{a}dx^{b}+R^{2}d\Omega^{2}_2~~~~~~~~~(1a)$$
where $$h_{ab}=diag\left(-1,
\frac{a^{2}}{1-kr^{2}}\right)~~~,~~~(a,~b=0,1~with~~x^{0}=t,
x^{1}=r)$$ is the metric on the plane normal to the spherical
surface of symmetry and
$$d\Omega^{2}_2=d\theta^{2}+sin^{2}\theta d\phi^{2}~is~ the~ metric~
on~ unit~ two~ sphere.$$ $R=ar$ is the radius of the
sphere(area-radius), $a(t)$ is the scale factor and $k=0, \pm1$
stands for flat, closed and open model of our universe
respectively. Let us denote by
$(\rho_e,p_e),~(\rho_m,p_m),~(\rho_r,p_r)$ the matter density and
thermodynamic pressure of dark energy, dark matter and radiation
respectively. Assuming barotropic equation of state for the
individual matter components we write
\begin{equation}
p_l=\omega_l \rho_l~,~~~~(l\equiv(e,m,r))
\end{equation}

The two Friedmann equations for the present model are

\begin{equation}
H^{2}+\frac{k}{a^{2}}=\frac{8\pi G}{3}\rho
\end{equation}
\begin{equation}
\dot{H}-\frac{k}{a^{2}}=-4\pi G\left(\rho+p\right)
\end{equation}

where $\rho=\rho_e+\rho_m+\rho_r$ is the total matter density and
the resulting thermodynamic pressure $p=p_e+p_m+p_r$. Due to
interaction among the matter constituents the energy conservation
relations are

\begin{equation}
\dot{\rho_{e}}+3H(1+\omega_{e}) \rho_{e}=-\Gamma
\end{equation}
$$
\dot{\rho_{m}}+3H(1+\omega_{m})
\rho_{m}=\Gamma'~~~~~~~~~~~~~~~~~~~~~~~~~~~~~~~~~~~~~~~~(5a)
$$
and
$$
\dot{\rho_{r}}+3H(1+\omega_{r})\rho_{r}=\Gamma-\Gamma'~~~~~~~~~~~~~~~~~~~~~~~~~~~~~~~~~~~~~(5b)
$$
Here the two interaction terms $\Gamma$ and $\Gamma'$ are in
general not constants (may have arbitrary forms) and distinct. The
sign of $\Gamma$ and $\Gamma'$ will indicate the direction of
matter flow. For example, if $\Gamma>0$ then there is an energy
flow from DE to the other two matter distribution while
$\Gamma'<0$ indicates an energy loss from the DM sector to the
other two constituents and so on. According to ref.[14] the
ansatzs for the interaction terms may be chosen as

\begin{equation}
\Gamma=\mu_e \rho_e~~and~~\Gamma'=\mu_e \rho_e
\end{equation}
and we have three non-interacting fluids with conservation
equations

$$
\dot{\rho_{e}}+3H(1+\omega_{e}^{eff}) \rho_{e}=0
$$
\begin{equation}
\dot{\rho_{m}}+3H(1+\omega_{m}^{eff})\rho_{m}=0
\end{equation}
and
$$
\dot{\rho_{r}}+3H(1+\omega_{r}^{eff})\rho_{r}=0
$$

Here the form of the modified state parameters are

\begin{equation}
\omega_D^{eff}=\omega_e+\frac{\mu_e}{3H} ~,~~~~~
\omega_m^{eff}=\omega_m-\frac{\mu_m}{3H}~~and~~~~\omega_r^{eff}=\omega_r-\frac{\mu_e
u}{3H}+\frac{\mu_m v}{3H}
\end{equation}
where $~u=\frac{\rho_e}{\rho_r}~,~v=\frac{\rho_m}{\rho_r}$ is the
ratio of two energy densities. Hence from (7) we have

$$
\dot{\rho}+3H(1+\omega) \rho=0~~~~~~~~~~~~~~~~~~~~~~~~(7a)
$$

where $$\omega= \alpha \omega_D + \beta \omega_m+ \delta
\omega_r~~~~~~~~~~~~~~~(8a)$$

with
$$\alpha=\frac{\rho_d}{\rho},~\beta=\frac{\rho_m}{\rho}~and~\delta=\frac{\rho_r}{\rho}.$$
Thus the interacting 3-components fluid distribution can be
considered as a single fluid with equation of state parameter
given by (8a).\\

\section{\normalsize\bf{Thermodynamics of the Universe bounded by the Event Horizon:}}

We start this section with the idea of horizons for the present
model. For the space-time metric given by equation (1a) the
apparent horizon $(R_A)$ is defined as

$$
h^{ab}\partial_{a}R\partial_{b}R=0
$$

or explicitly it gives

\begin{equation}
R_A=\frac{1}{\sqrt{H^2+\frac{k}{a^2}}}
\end{equation}

One may note that this apparent horizon coincides with the
trapping horizon $[15]$ and for flat case apparent horizon and
Hubble horizon $(R_H=\frac{1}H)$ coincides. The radius of the
event horizon $(R_E)$ is mathematically given by

\begin{equation}
R_{E}=a\int^{\infty}_{t}\frac{dt}{a}=a\int^{\infty}_{a}\frac{da}{Ha^{2}}
\end{equation}
It is to be noted that event horizon exists only for accelerating
universe. In terms of the conformal time $\tau~[16]$

\begin{equation}
\tau=-\int_{t}^{\infty}\frac{dt}{a(t)}~~~~~~~~~~~~~~~~~~~~~~~~~~~~~~|\tau|<\infty
\end{equation}

the event horizon can be written as [16]

$$R_{E}=-a ~sinh(\tau)~~~~~~~~~~~~~~k=-1$$

\begin{equation}
R_{E}=-a\tau~~~~~~~~~~~~~~~~~~~~~k=0
\end{equation}

$$R_{E}=-a~sin(\tau)~~~~~~~~~~~~~~~~~k=+1$$

Note that if $|\tau |=\infty$, event horizon does not exist.\\

The change of different horizon radii with the evolution of the
universe are given by

\begin{equation}
\dot{R}_H=-\frac{\dot{H}}{H^2}
\end{equation}

\begin{equation}
\dot{R}_{A}=-H\left(\dot{H}-\frac{k}{a^{2}}\right)R_{A}^{3}
\end{equation}

\begin{equation}
\dot{R}_{E}=H R_{E}-\sqrt{1-\frac{k}{a^{2}}R_{E}^{2}}
\end{equation}

We see that radius of the event horizon increases through out the
evolution of the universe so long as $R_E>R_A$ and it does not
depend on the nature of the matter in the universe. On the other
hand both $R_A$ and $R_H$ increases with the evolution of the
universe in the quintessence era but decreases in the phantom era.
(For detailed discussion see ref[17]).\\

To find the entropy variation of the surface of the event horizon
we start with unified first law

\begin{equation}
dE=A\psi+WdV
\end{equation}

where,
\begin{equation}
E=\frac{R}{2G}\left(1-h^{ab}\partial_{a}R\partial_{b}R\right)
\end{equation}
is the Misner-sharp energy.
$$
\Psi=\psi_{a}dx^{a},~~~~is~ the~ energy ~supply $$
\begin{equation}\psi_a=T_a^b
\partial_b R+  \partial_a R,~~~~is~ the ~energy~ flux
\end{equation}
 and $$W=-\frac{1}{2}Trace (T),~~  is~ the~
work~ function~ $$

For the present model we have

$$A \Psi= 2 \pi R^2 (\rho+p)[-2HRdt+dR]$$
\begin{equation}WdV=2 \pi R^2 (\rho-p)dR
\end{equation}

Hence from the Clausius relation on the event horizon
$$T_{E}dS_{E}=dQ=-dE=4\pi R_{E}^{3}H\left(\rho+p\right)dt$$
i.e.,
\begin{equation}
T_{E}\frac{dS_{E}}{dt}=4\pi R_{E}^{3}H\left(\rho+p\right)
\end{equation}
Where $S_{E}$ and $T_{E}$ are respectively the entropy and the
temperature on the event horizon. For equilibrium thermodynamics
we assume $T_{E}$ as the temperature of the matter inside the
event horizon. For variation of the entropy ($S_{I}$) of the
matter distribution we take help of the Gibb's equation
\begin{equation}
T_{E}dS_{I}=dE_{I}+pdV_{I}
\end{equation}
where $V_{I}=\frac{4}{3}\pi R_{E}^{3}$, the volume bounded by the
event horizon and $E_{I}=\rho . V_{I}$. Thus using the combined
energy conservation relation (7a) and the variation of the radius
of the event horizon , i.e., eq. (15) we obtain
\begin{equation}
T_{E}\frac{dS_{I}}{dt}=-4\pi
R_{E}^{2}\left(\rho+p\right)\sqrt{1-\frac{k}{a^{2}}R_{E}^{2}}
\end{equation}
Hence combining (20) and (22) we obtain
\begin{equation}
T_{E}\frac{d}{dt}\left(S_{E}+S_{I}\right)=4\pi
R_{E}^{2}\left(\rho+p\right)\left[HR_{E}-\sqrt{1-\frac{k}{a^{2}}R_{E}^{2}}\right]
\end{equation}

From the above thermodynamical analysis the conclusions are the
following :\\

$(I)$ The time variation of the entropy of the horizon and that of
the matter distribution inside the horizon are of opposite
character. In the quintessence era (i.e., when the resulting
matter satisfies week energy condition) $S_{E}$ is an increasing
function while $S_{I}$ decreases with the evolution. However in
the phantom era the entropy functions exchange their role, i.e.,
entropy of the horizon decreases while entropy of the matter
distribution increases with the evolution.\\

$(II)$ The validity of the generalized second law of
thermodynamics (GSLT) depends both on the nature of the matter and
on the evolution of the horizons, i.e., GSLT will hold if\\

$(a)$ $\rho+p>0$ and $\dot{R}_{E}>0$ ,i.e., $R_{E}>R_{A}$\\

$(b)$ $\rho+p<0$ and $\dot{R}_{E}<0$ ,i.e., $R_{E}<R_{A}$\\

The first possibility indicates that the resulting matter should
be of quintesence nature , i.e.,
$$1+\omega>0$$
$$1+\alpha \omega_{d}+\beta \omega_{m}+\delta\omega_{r}>0$$
\begin{equation}
1+\frac{\Omega_{d}\omega_{d}+\Omega_{m}\omega_{m}+\Omega_{r}\omega_{r}}{1-\Omega_{k}}>0
\end{equation}
For the other possibility the restrictions are
\begin{equation}
1-\Omega_{k}+\Omega_{d}\omega_{d}+\Omega_{m}\omega_{m}+\Omega_{r}\omega_{r}<0
\end{equation}
and $R_{E}<R_{A}$.\\

$(III)$ The validity of GSLT doe not depend on the interaction
terms, it only depends on the equation of state parameter for each
component.\\

{\bf References:}\\
\\
$[1]$ Riess A. G., et al., \it{AstroPhys J.} {\bf 607 } (2004)
665.\\\\
$[2]$ C. L. Bennett et al., \it{Astrophys. J. Suppl.} {\bf 148},
(2003),1 .\\\\
$[3]$ M. Tegmark et al. [SDSS Collaboration], \it{Phys. Rev. D},
{\bf 69}, (2004), 103501 . \\\\
$[4]$ S. W. Allen, et al., \it{Mon. Not. Roy. Astron. Soc.}, {\bf
353}, (2004), 457 .\\\\
$[5]$  P. J. Steinhardt, it{Critical Problems in Physics} (1997),
Princeton University Press.\\\\
$[6]$ J. Sola and H. Stefancic, \it{Phys. Lett. B} {\bf
624},(2005) 147 ; J. Sola and H. Stefancic, \it{Mod. Phys. Lett.
A} {\bf 21}, (2006) 479; I. L.
Shapiro and J. Sola, \it{Phys. Lett. B} {\bf 682}, (2009) 105 .\\\\
$[7]$  B. Ratra and P. J. E. Peebles, Phys. Rev. D 37, 3406
(1988); C. Wetterich, Nucl. Phys. B 302, 668 (1988); A. R. Liddle
and R. J. Scherrer, Phys. Rev. D 59, 023509 (1999); I. Zlatev, L.
M. Wang and P. J. Steinhardt, Phys. Rev. Lett. 82, 896 (1999); Z.
K. Guo, N. Ohta and Y. Z. Zhang, Mod. Phys. Lett. A 22, 883
(2007); S. Dutta, E. N. Saridakis and R. J. Scherrer, Phys. Rev. D
79, 103005 (2009).\\\\
$[8]$ R. R. Caldwell, \it{Phys. Lett. B} {\bf 545}, (2002) 23 ; R.
R. Caldwell, M. Kamionkowski and N. N. Weinberg, \it{Phys. Rev.
Lett.} {\bf 91}, (2003) 071301 ; S. Nojiri and S. D. Odintsov,
\it{Phys. Lett. B} {\bf 562},(2003) 147 ; V. K. Onemli and R. P.
Woodard, \it{Phys. Rev. D} {\bf 70}, (2004) 107301 ; M. R. Setare,
J. Sadeghi, A. R. Amani, \it{Phys. Lett. B} {\bf 666}, (2008) 288;
M. R. Setare and E. N. Saridakis, \it{JCAP} {\bf 0903},(2009)
002 ; E. N. Saridakis, \it{Nucl. Phys. B} {\bf 819}, (2009) 6 116 .\\\\
$[9]$ B. Feng, X. L. Wang and X. M. Zhang, \it{Phys. Lett. B} {\bf
607}, 35 (2005); Z. K. Guo, et al., \it{Phys. Lett. B} {\bf 608},
(2005) 177 ; M.-Z Li, B. Feng, X.-M Zhang, \it{JCAP}, {\bf
0512},(2005) 002 ; B. Feng, M. Li, Y.-S. Piao and X. Zhang,\it{
Phys. Lett. B} {\bf 634}, (2006) 101 ; M. R. Setare, \it{Phys.
Lett. B} {\bf 641},(2006) 130 ; W. Zhao and Y. Zhang, \it{Phys.
Rev. D} {\bf 73},(2006) 123509 ; M. R. Setare, J. Sadeghi, and A.
R. Amani, \it{Phys. Lett. B} {\bf 660}, (2008) 299 ; M. R. Setare
and E. N. Saridakis, \it{Phys. Lett. B} {\bf 66}8, (2008) 177 ; M.
R. Setare and E. N. Saridakis, \it{JCAP} {\bf 0809},(2008) 026 ;
M. R. Setare and E. N. Saridakis, \it{Int. J. Mod. Phys. D} {\bf
18},(2009) 549 ; Yi-Fu Cai, Emmanuel N. Saridakis, Mohammad R.
Setare, Jun-Qing Xia, arXiv:0909.2776 [hep-th]\\\\
$[10]$ S. D. H. Hsu :- {\it Phys. Lett. B} {\bf 594}, 13 (2004);
{\it M. Li, Phys. Lett. B} {\bf 603}, 1 (2004); Q. G. Huang and M.
Li, {\it JCAP} {\bf 08} (2004) 013; {\it M. Ito, Europhys. Lett.}
{\bf 71}, 712 (2005); X. Zhang and F. Q. Wu :- {\it Phys. Rev. D}
{\bf 72}, 043524 (2005); D. Pavon and W. Zimdahl :- {\it Phys.
Lett. B} {\bf 628}, 206 (2005); S. Nojiri and S. D. Odintsov :-
{\it Phys. Rev. D} {\bf 81}, 023007 (2010) 023007-5. {\it Relativ.
Gravit.} {\bf 38}, 1285 (2006); E. Elizalde, S. Nojiri, S. D.
Odintsov, and P. Wang :- {\it Phys. Rev. D} {\bf 71}, 103504
(2005); H. Li, Z. K. Guo, and Y. Z. Zhang:- {\it IJMPD} {\bf 15},
869 (2006); E. N. Saridakis:- {\it  Phys. Lett. B} {\bf 660}, 138
(2008); {\it JCAP} {\bf 04}, 020(2008) ;
{\it Phys. Lett. B} {\bf 661}, 335 (2008).\\\\
$[11]$ P. Horava, {\it Phys. Rev. D} {\bf 79}, 084008 (2009); G.
Calcagni, arXiv:0904.0829 [hep-th]; E. Kiritsis and G. Kofinas,
{\it Nucl. Phys. B} {\bf 821}, 467 (2009); H. Lu, J. Mei and C. N.
Pope, arXiv:0904.1595 [hep-th]; C. Charmousis, G. Niz, A. Padilla
and P. M. Saffin, arXiv:0905.2579 [hep-th]; E. N. Saridakis,
arXiv:0905.3532 [hep-th]; X. Gao, Y. Wang, R. Brandenberger and A.
Riotto, arXiv:0905.3821 [hep-th]; M. i. Park, arXiv:0905.4480
[hep-th]; Y. F. Cai and E. N. Saridakis, arXiv:0906.1789 [hep-th];
M. Botta-Cantcheff, N. Grandi and M. Sturla, arXiv:0906.0582
[hep-th]; M. R. Setare, arXiv:0909.0456 [hep-th]; C. Germani, A.
Kehagias and K. Sfetsos, {\it JHEP} {\bf 0909}, 060 (2009); G.
Leon and E. N. Saridakis, {\it JCAP} {\bf 0911}, 006 (2009);
 Mubasher Jamil, Emmanuel N. Saridakis, M. R.
Setare, arXiv:1003.0876[gr-qc].\\\\
$[13]$ J.D.Bekenstein, \it{Phys. Rev. D} {\bf 7} 2333 (1973).\\\\
$[14]$ M.R. Setare , {\it JCAP} {\bf 01} 023 (2007);Mubasher
Jamil, Emmanuel N. Saridakis, M. R. Setare {\it
Phys.Rev.D} {\bf 81} 023007, (2010).\\\\
$[15]$ Subenoy Chakraborty, Ritabrata Biswas and Nairwita
Mazumder, {\it arXiv:1006.1169 [gr-qc] }.\\\\
$[16]$ P.C.W. Davis , {\it Class. Quantum Grav.} {\bf 5} (1998)
1349.\\\\
$[17]$ Subenoy Chakraborty, Ritabrata Biswas and Nairwita
Mazumder, {\it arXiv:1006.0881 [gr-qc] }.\\\\

\end{document}